\documentstyle[epsfig]{mn}
\begin{document}
\LARGE
\normalsize

\title[GRS\,1915+105]
{The variable radio emission from GRS\,1915+105}

\author[G.~G.~Pooley \& R.~P.~Fender]
{G.~G.~Pooley$^1$ and R.~P.~Fender$^2$\\
$^1$Mullard Radio Astronomy Observatory, Cavendish Laboratory,
Madingley Road, Cambridge CB3 0HE\\
$^2$Astronomy Centre, University of Sussex, Falmer, Brighton BN1 9QH\\
}

\maketitle

\begin{abstract}
We present data on the monitoring of the Galactic X-ray transient
GRS\,1915+105 at 15 GHz with the Ryle Telescope. We have found
quasi-periodic oscillations with
periods in the range 20--40 min which are tentatively associated with
the soft-X-ray variations on the same time-scale. The overall behaviour of
the radio emission is shown to vary in a strong association with the
X-ray emission as recorded by the {\it RXTE} all-sky monitor.

\end{abstract}


\section{Introduction}

Castro-Tirado et al.\ (1992) reported the discovery of the hard-X-ray transient
source GRS\,1915+105, using the WATCH instrument on the {\it GRANAT} satellite. Its
X-ray emission at both high and low energies has proved to have a rich
structure -- see, e.g., Paciesas et al.\ (1996) and Greiner et al.\ (1996). In the
radio regime, the source is no less remarkable; Mirabel \& Rodr\'\i guez (1994)
discovered a double-sided relativistic ejection of radio-emitting material.
Adopting a model based on symmetrical ejection, they derive an angle to the
line of sight of 70 degrees and a velocity of ejection of 0.92$c$. The distance,
from 21-cm H\,{\sc i} absorption, is estimated to be 12.5 kpc, consistent with the
relativistic-expansion model. Rodr\'\i guez et al.\ (1995) and Foster et al.\ (1996)
have presented flux-density monitoring data at a range of frequencies.

From this monitoring, it was apparent that the flux density in the radio regime varies on many
time-scales, and we started a systematic monitoring program at 15 GHz with the
Ryle Telescope in 1995 August. One surprising feature, apparently periodic
oscillations with periods in the range 20--40 min, has been reported in two
IAU Circulars (Pooley 1995, 1996). This phenomenon, and other patterns of
variation including the relationship to the {\it RXTE} monitoring data, are considered
in more detail in this paper.

\section{Observations}

\begin{figure*}
\centering
\leavevmode\epsfig{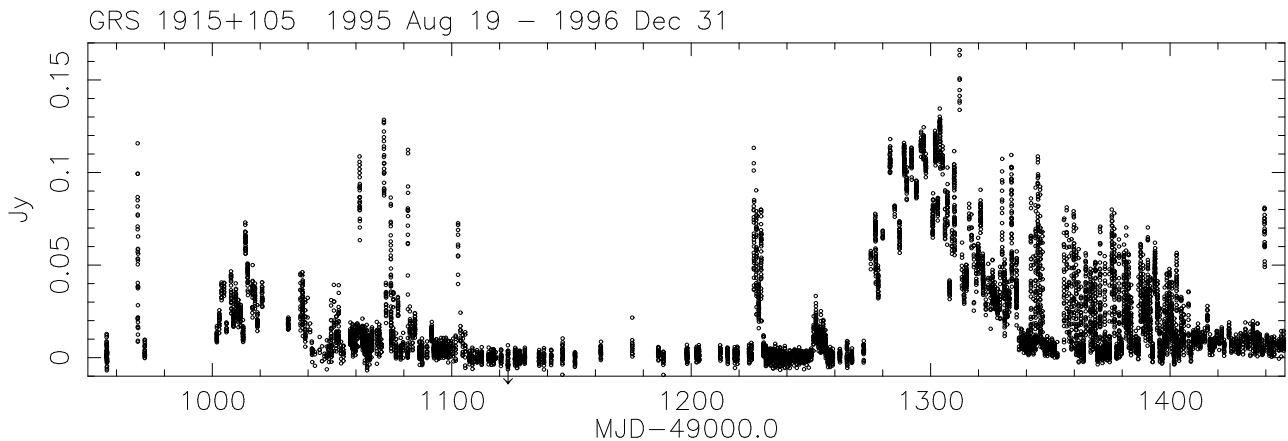}
\caption{Flux density of GRS\,1915+105 at 15 GHz from 1995 Aug 10 to
1996 Dec 31. Each point plotted represents an integration of 5 min.}
\end{figure*}

The Ryle Telescope (RT: Jones 1991), the upgraded Cambridge 5-km
Telescope, is used primarily for observations related to
microwave-background studies. It is an E--W synthesis telescope for
which the majority of observations have durations of 12 h. In the
inevitable gaps in that program, projects such as monitoring of
variable sources are often carried out. All observations reported here
are centred on 15.2 GHz with a bandwidth of 350 MHz. The
linearly-polarized feeds are sensitive to Stokes I+Q; no attempt was
made to measure polarization properties.

Since much of the work of the telscope is at modest resolutions, the four
mobile aerials are usually set in a compact array configuration within 100~m of
the nearest fixed aerial; the resulting 10 baselines give a resolution of some
30 arcsec at 15 GHz when mapping. Atmospheric phase fluctuations are not a
serious problem on these baselines. Most of the data reported in this
paper are derived from the compact array and not the longer baselines which
were also available. One difficulty arises in the compact array from the
declination of this particular source, in that some antennas become severely
shadowed at extreme hour angles. In practice only hour angles within 3h of the
meridian were normally used to avoid shadowing. The observations were
interleaved with those of a calibrator, B1920+154, using a 15 + 2.5 minute
cycle, so that the instrumental phase variations can be determined and removed. The
flux-density scale is set by nearby observations of 3C48 or 3C286, and is
believed to conform to the scale of Baars et al.\ (1977). The measured flux
density of B1920+154 was used as a check that the system was working well; they
show that the day-to-day calibration of the flux density scale is consistent
within 3 percent rms.
Data are normally integrated over 32-second (sidereal)
intervals, although several observations were made using 8 seconds in order to search for
faster variations. The correctly-phased baselines are added to produce,
effectively, a phased-array response; since the position of the source is
adequately known, the in-phase component then represents an unbiased estimate
of its flux density. The typical rms noise on a single 32-second sample when
observing in this mode is 6 mJy. Since the estimates are unbiased, the noise
level reduces as the square root of the integration time.

Rodr\'\i guez et al.\ (1995) show a map of nearby sources, including the H\,{\sc ii}
 region G45.46+0.06 which has a total flux density of about 5 Jy at
this frequency. It lies about 12 arcmin from GRS\,1915+105 (compared
with a full-width to half-power primary beam of 6 arcmin). Using the
RT with the pointing centre on GRS\,1915+105 an image of the region
shows a response of some 3 mJy at the position of the H II region.
Since GRS\,1915+105 is variable and the observations are in any case
usually too short for satisfactory mapping,
there is a very small additional uncertainty in the flux densities when observing in the
`phased array' mode; values below about 1 mJy may be
unreliable. Some observations were made in a slightly more extended
array, when this problem is not significant.

\section{Results}

\subsection{The overall picture}

The data presented here run from 1995 Aug 10 to 1996 Dec 31. Individual
observations were of varying duration, typically between 1h and 6h. During
intervals of pronounced activity it was often possible to observe every day.
Fig. 1 shows some 9000 points, each one being a 5-min integration, over the the
whole of this time. The overall pattern of variations is apparent from this
plot, although the details are not.

Fig. 2 shows a series of individual observations, illustrating the
range of behaviours observed.
Particular features include:

1. Smoothly-varying flux density during major flares with decay times
of hours or days.  The flare starting near MJD 50275 (1996 July) was
characterised by smooth variations of flux density until it had almost
disappeared, at which time the emission became much more erratic.

2. Quasi-periodic oscillations (QPOs) and isolated short flare events, a
selection of which is shown in Fig. 2; these are discussed further
below.

3. Very low flux densities between active periods (e.g. MJD 50105 -- 50220,
Fig. 1).

\subsubsection{Quasi-periodic oscillations}

These remarkable features were first observed in late 1995, and reported in
IAU Circulars (Pooley 1995, 1996). One other example has also been
reported by Rodr\'\i guez \& Mirabel (1997).

We note the following features:

1. The `periods' vary in the range 20 -- 40 min. The most
frequently-observed periods are close to 40 min and 25 min; the event
reported by Rodr\'\i guez \& Mirabel (1997) had a period of 30 min. There are
clear instances when a change in the period occurs during an
observation (e.g. 1996 May 26, 1996 Sep 18), and there are also
instances when the gap between the maxima is erratic.
Isolated peaks
can be characterised by a rise-time close to 5 min and a decay which
is approximately exponential with a time-constant between
12 and 25 min. When the peaks are close together, they appear as
quasi-sinusuoidal variations, although it is often observed that the
rise is more rapid than the fall of each cycle.  We suggest that the
events themselves are similar, and they are triggered by releases of
energy on some short time-scale.  The amplitudes of the individual
peaks seldom exceed 100 mJy (the maximum flux density recorded in
the whole of this dataset is 170 mJy). Individual sequences of
oscillations often have nearly constant amplitudes and may have
high fractional modulations (see, e.g., 1996 Oct 12).

\begin{figure*}
\centerline{\epsfig{file=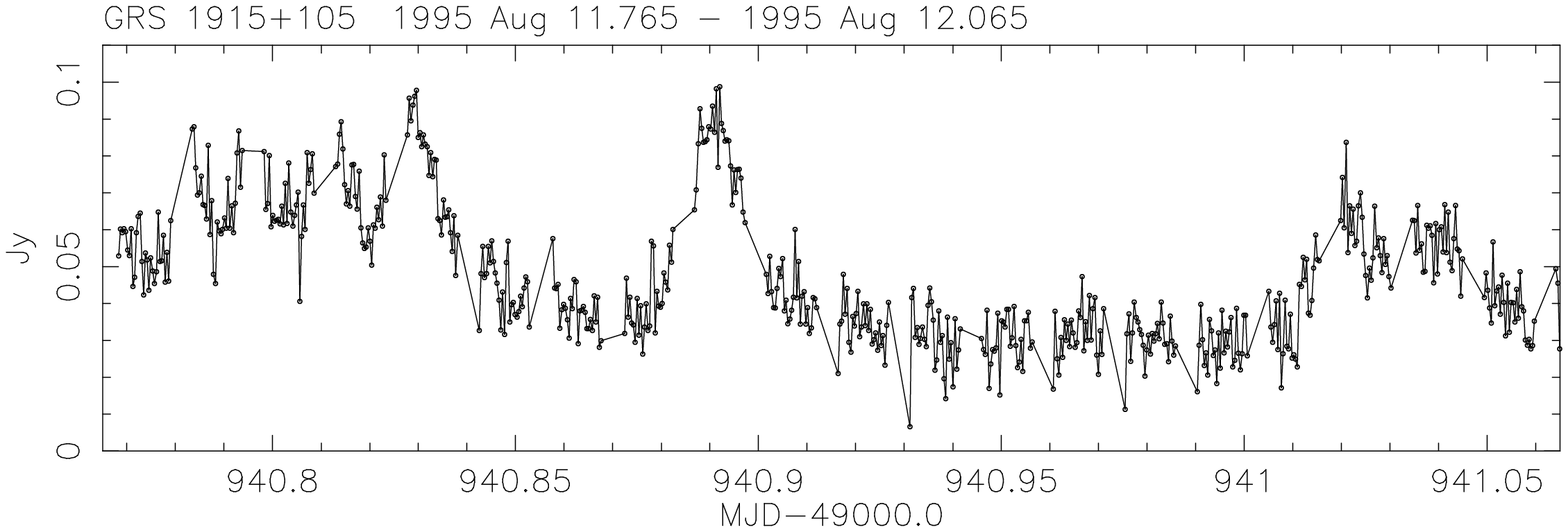,angle=270,width=85mm,clip=}\quad\epsfig{file=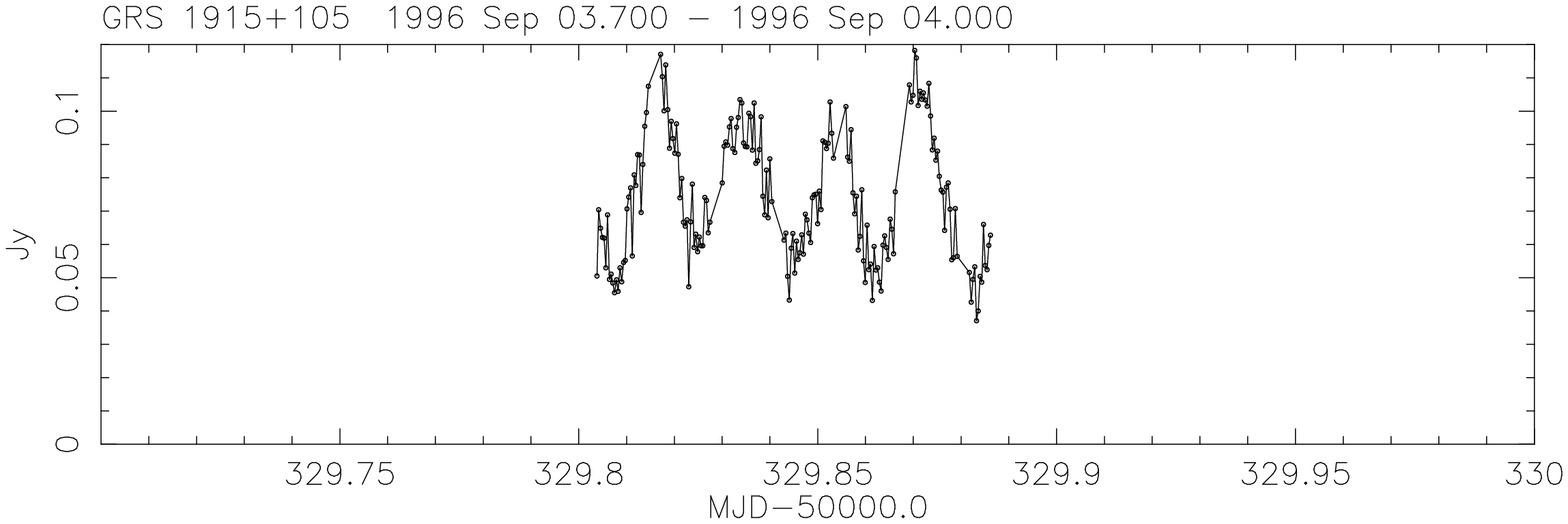,angle=270,width=85mm,clip=}}
\vspace{1mm}
\centerline{\epsfig{file=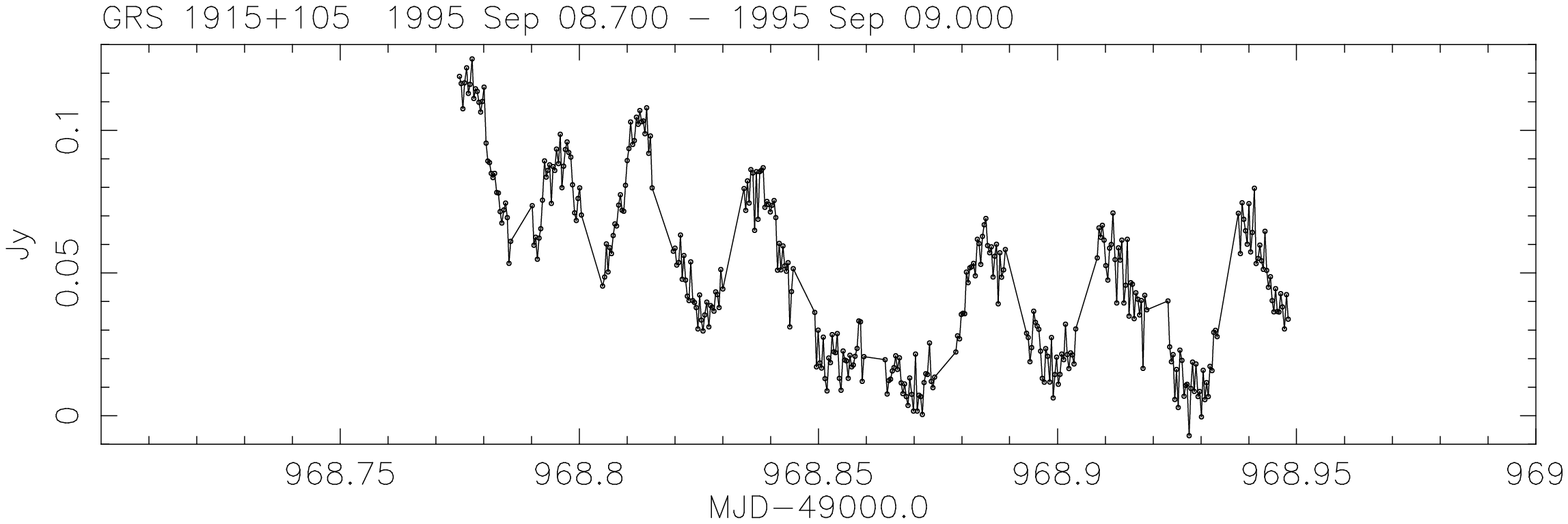,angle=270,width=85mm,clip=}\quad\epsfig{file=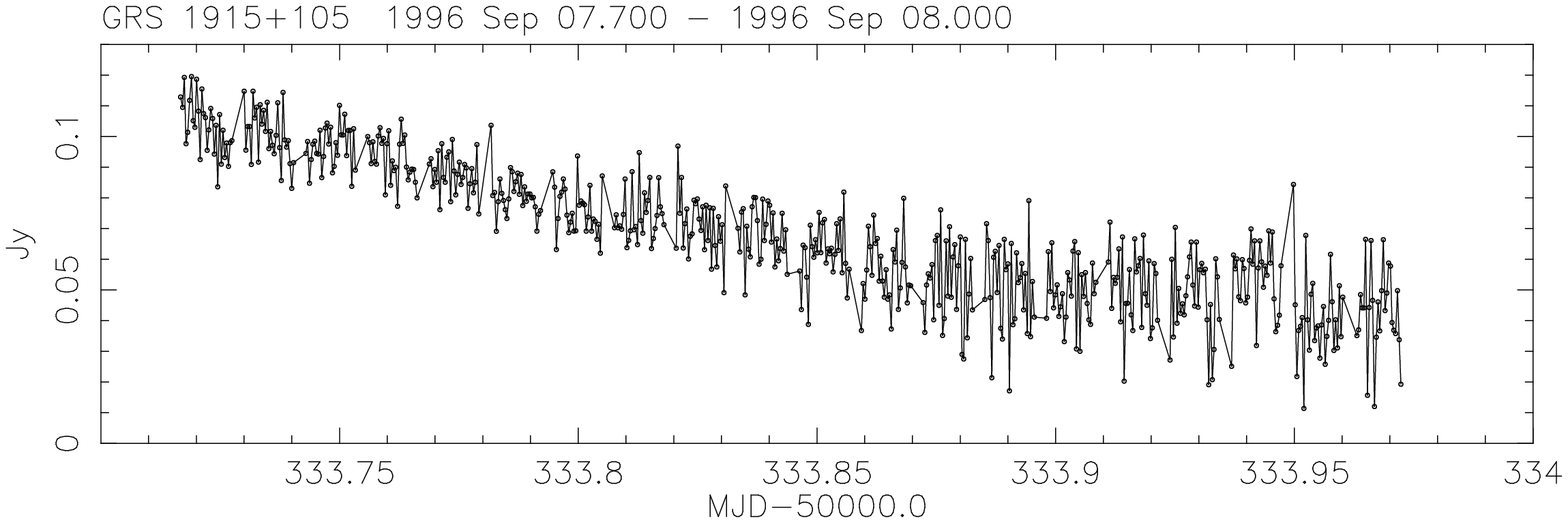,angle=270,width=85mm,clip=}}
\vspace{1mm}
\centerline{\epsfig{file=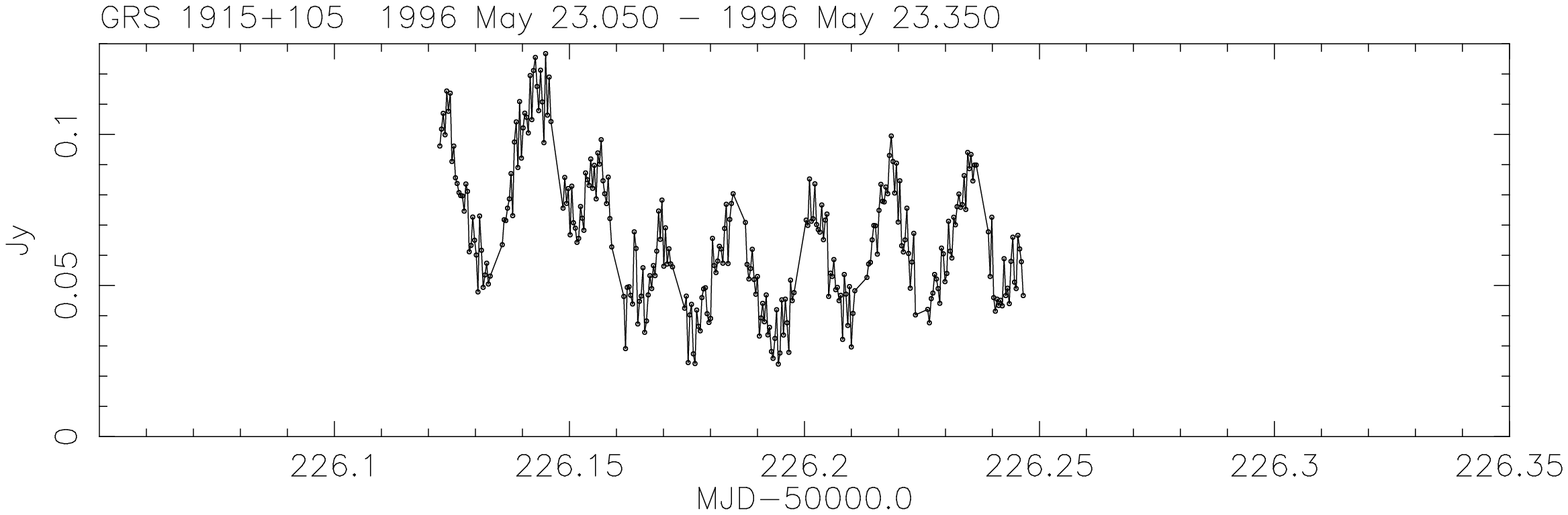,angle=270,width=85mm,clip=}\quad\epsfig{file=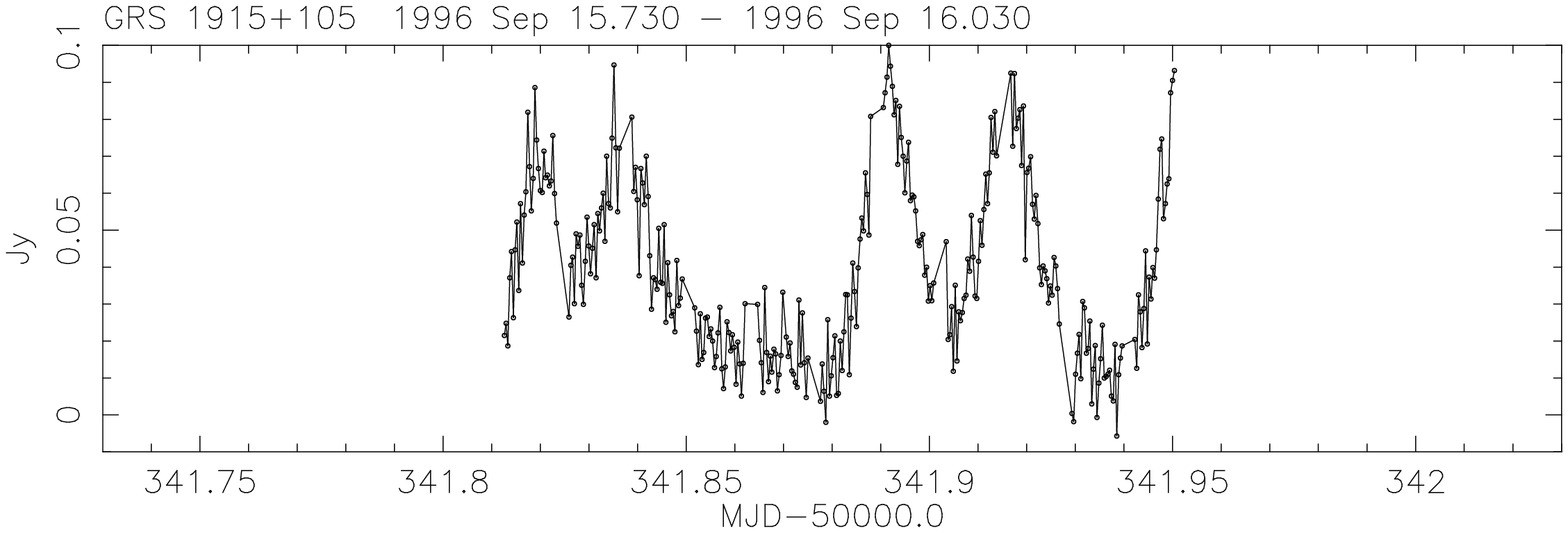,angle=270,width=85mm,clip=}}
\vspace{1mm}
\centerline{\epsfig{file=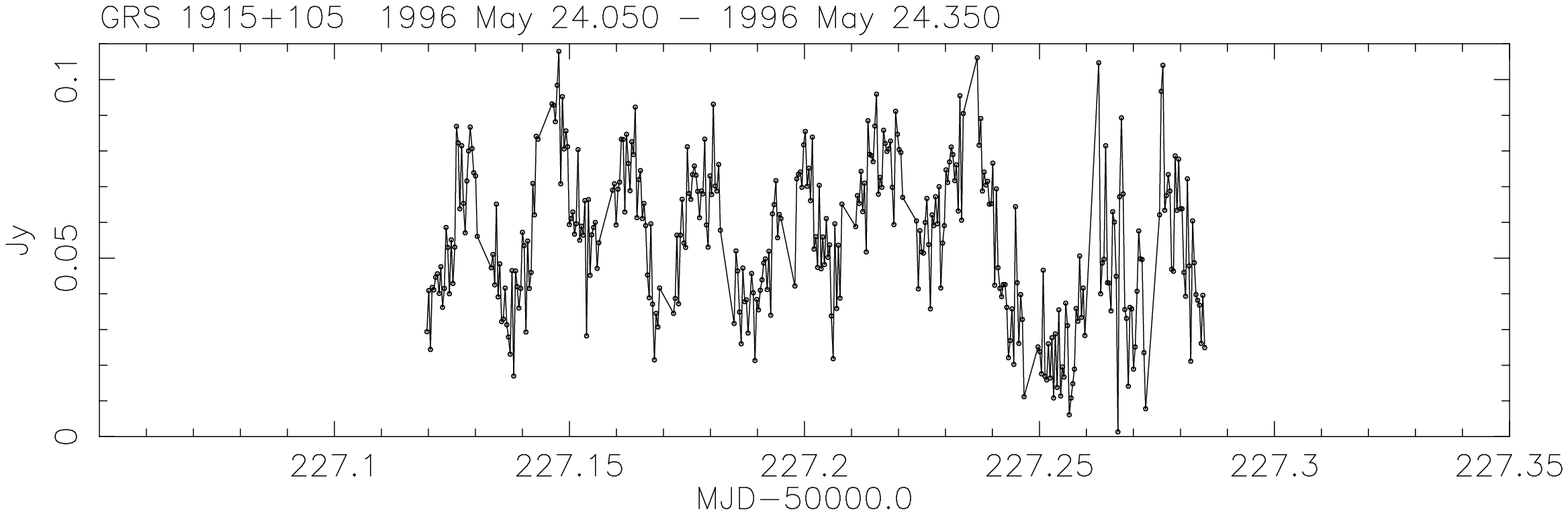,angle=270,width=85mm,clip=}\quad\epsfig{file=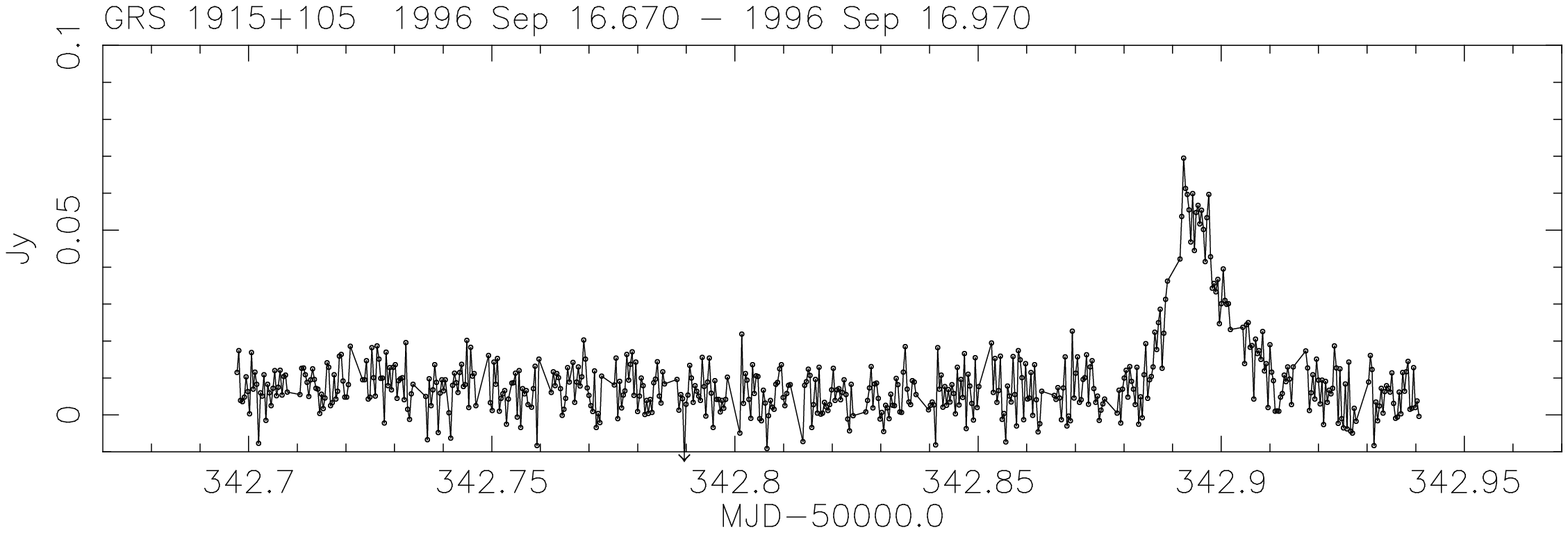,angle=270,width=85mm,clip=}}
\vspace{1mm}
\centerline{\epsfig{file=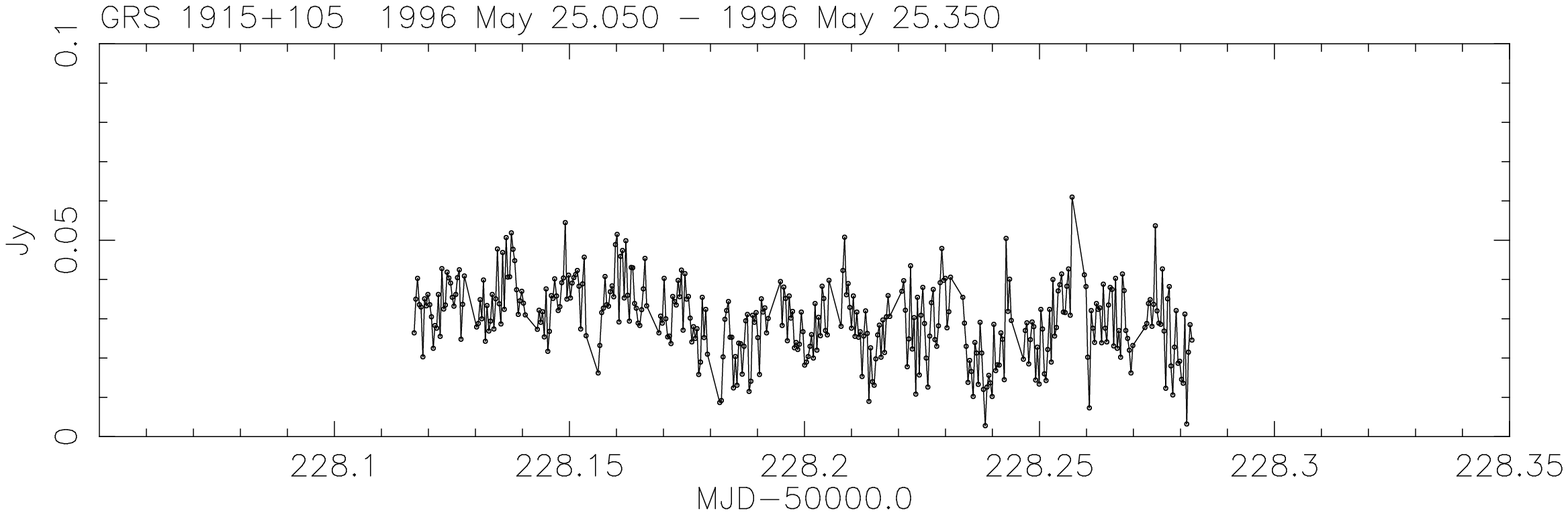,angle=270,width=85mm,clip=}\quad\epsfig{file=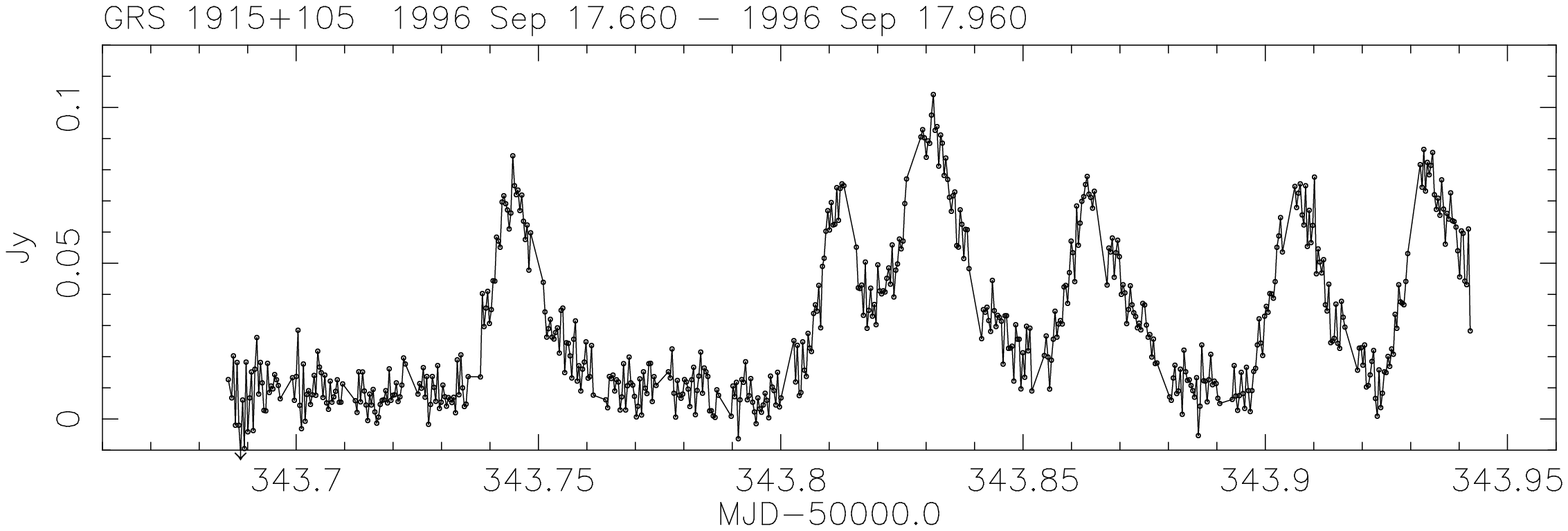,angle=270,width=85mm,clip=}}
\vspace{1mm}
\centerline{\epsfig{file=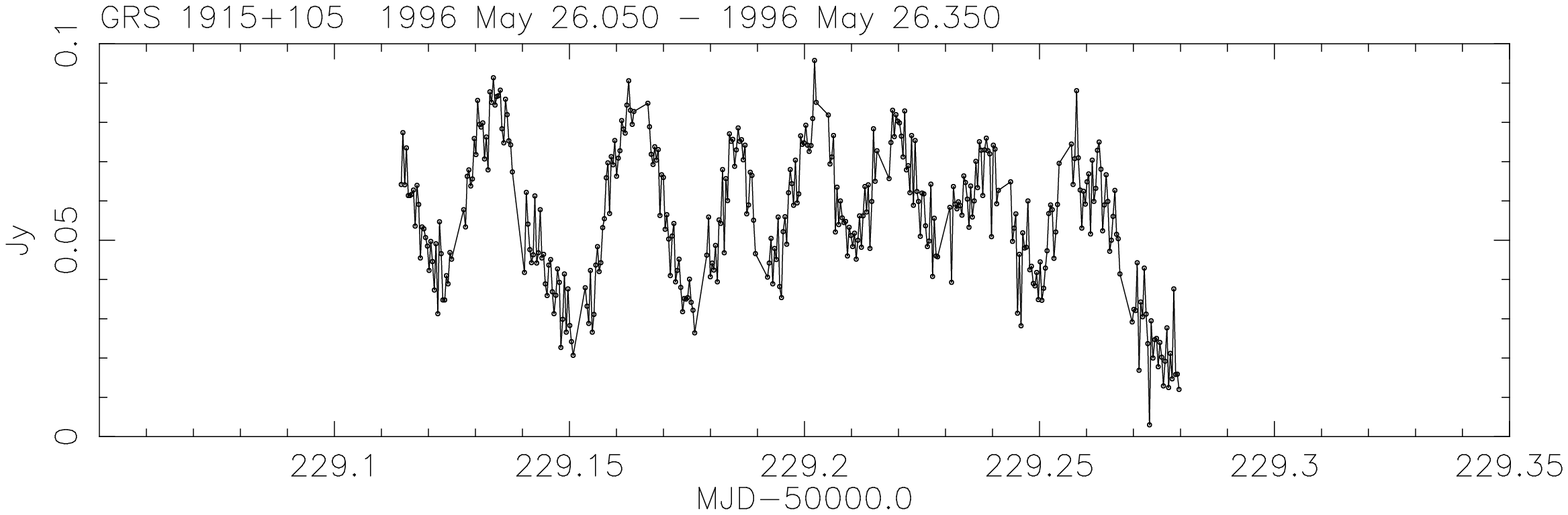,angle=270,width=85mm,clip=}\quad\epsfig{file=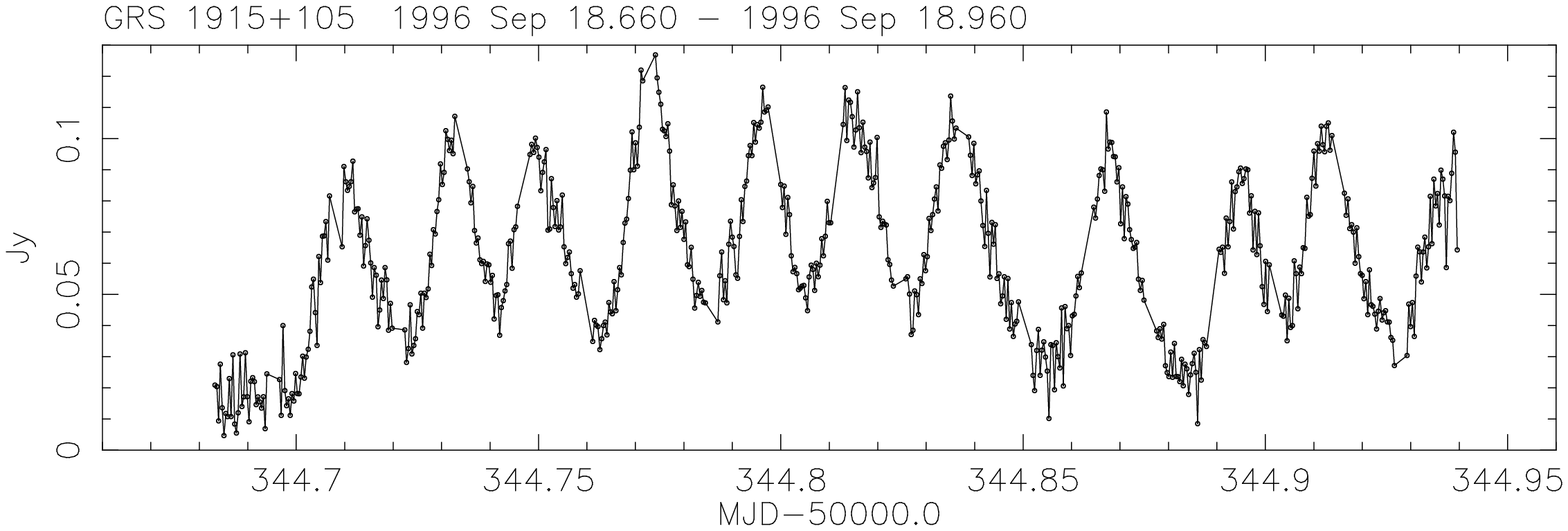,angle=270,width=85mm,clip=}}
\vspace{1mm}
\centerline{\epsfig{file=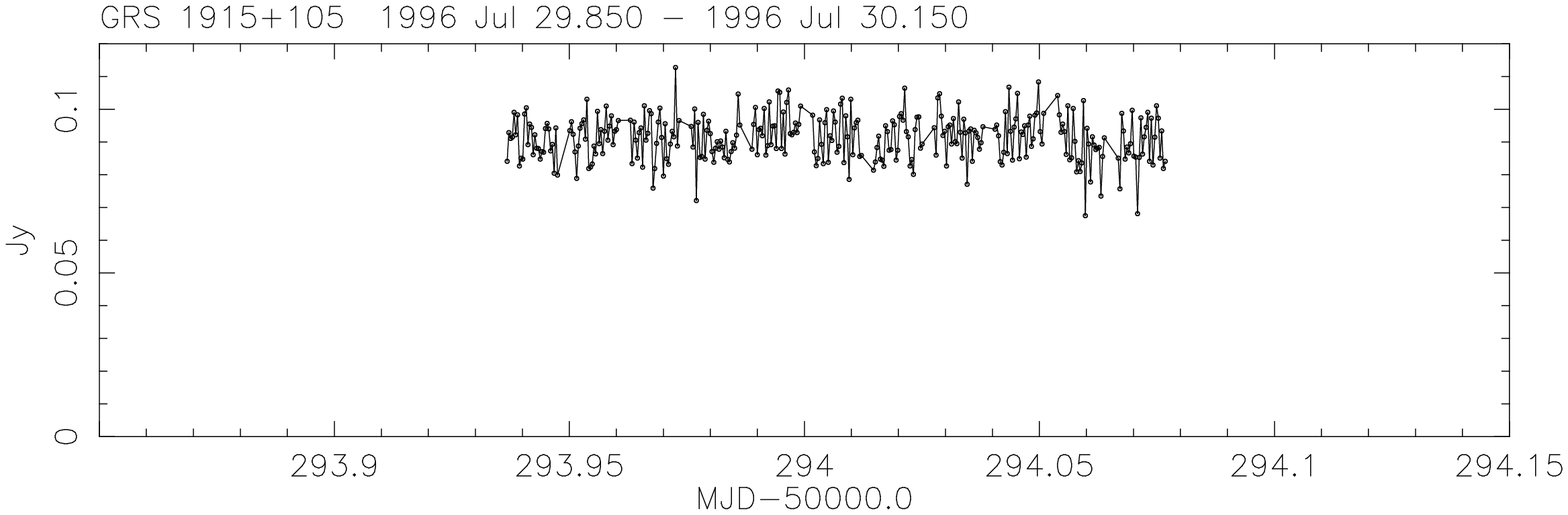,angle=270,width=85mm,clip=}\quad\epsfig{file=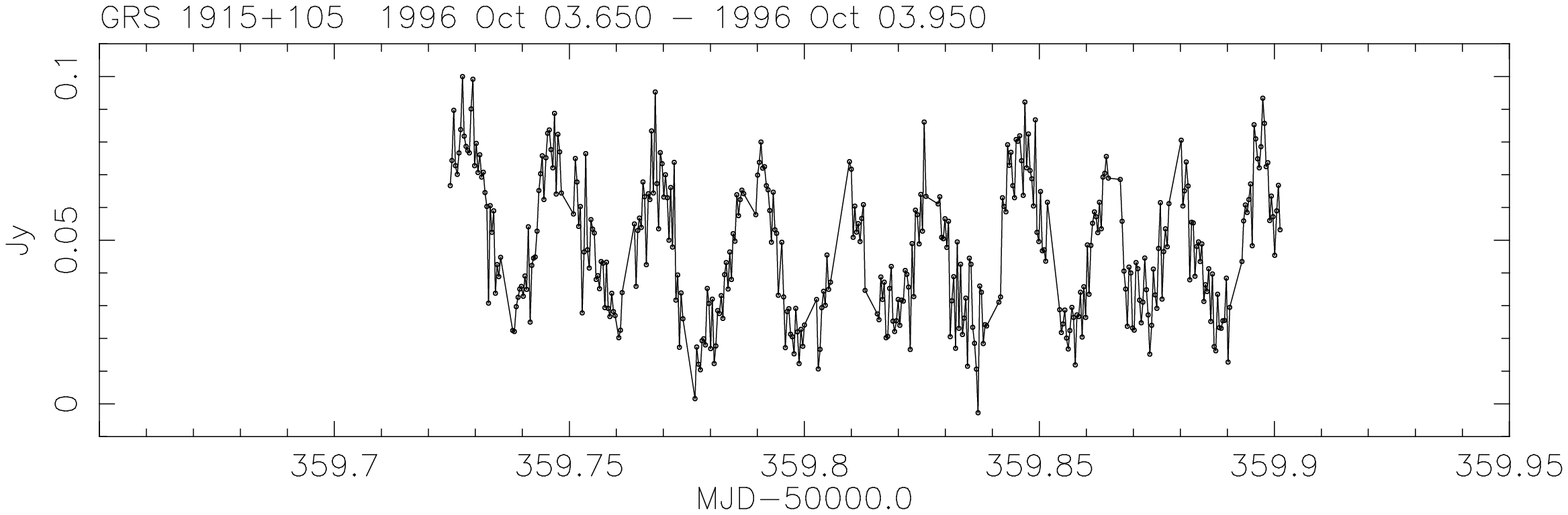,angle=270,width=85mm,clip=}}
\vspace{1mm}
\centerline{\epsfig{file=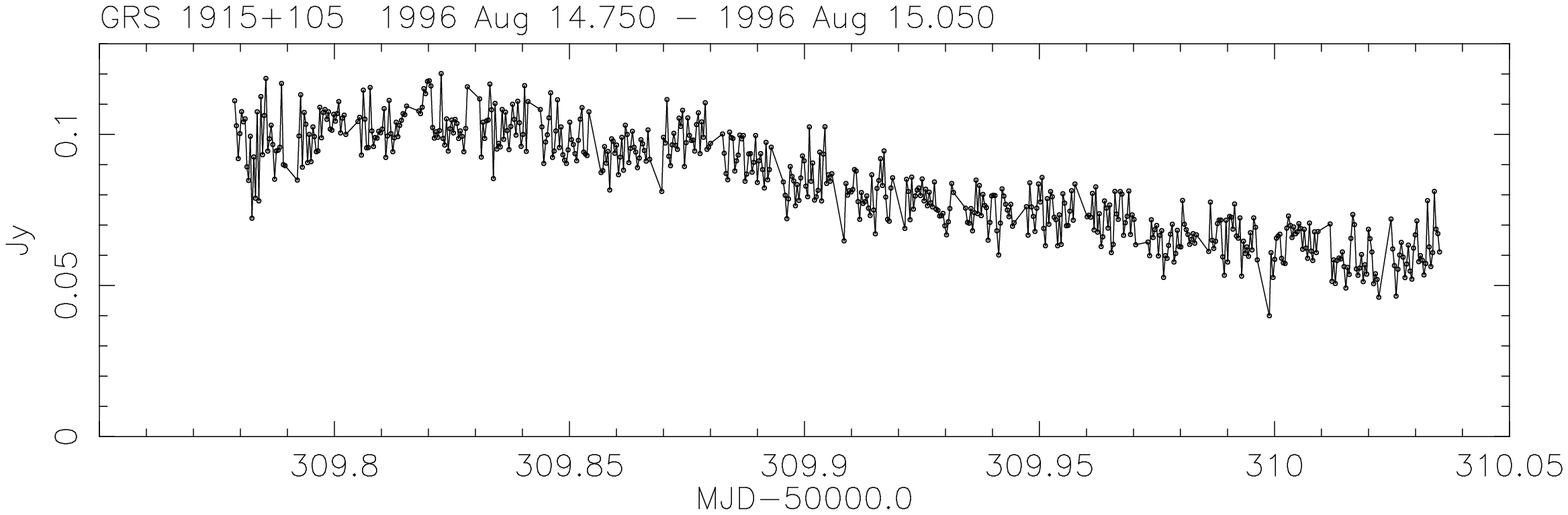,angle=270,width=85mm,clip=}\quad\epsfig{file=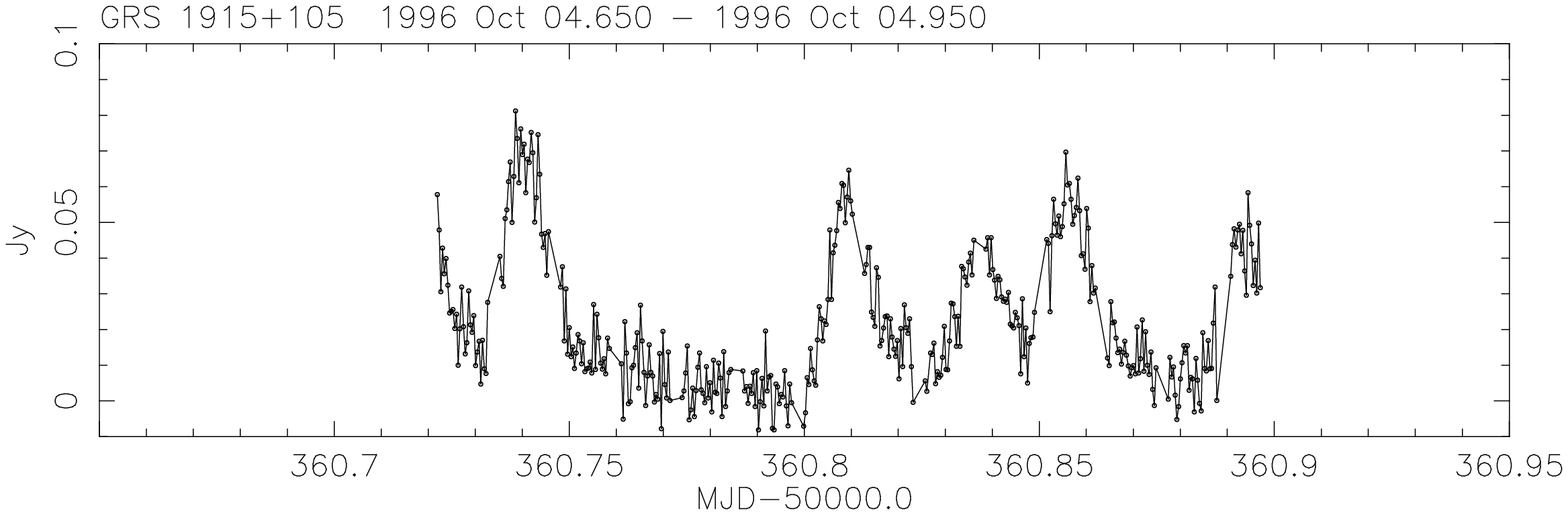,angle=270,width=85mm,clip=}}
\caption{Short-timescale behaviour of GRS\,1915+105 at 15 GHz; the integration time is 32 s}
\end{figure*}
\addtocounter{figure}{-1}
\begin{figure*}
\centerline{\epsfig{file=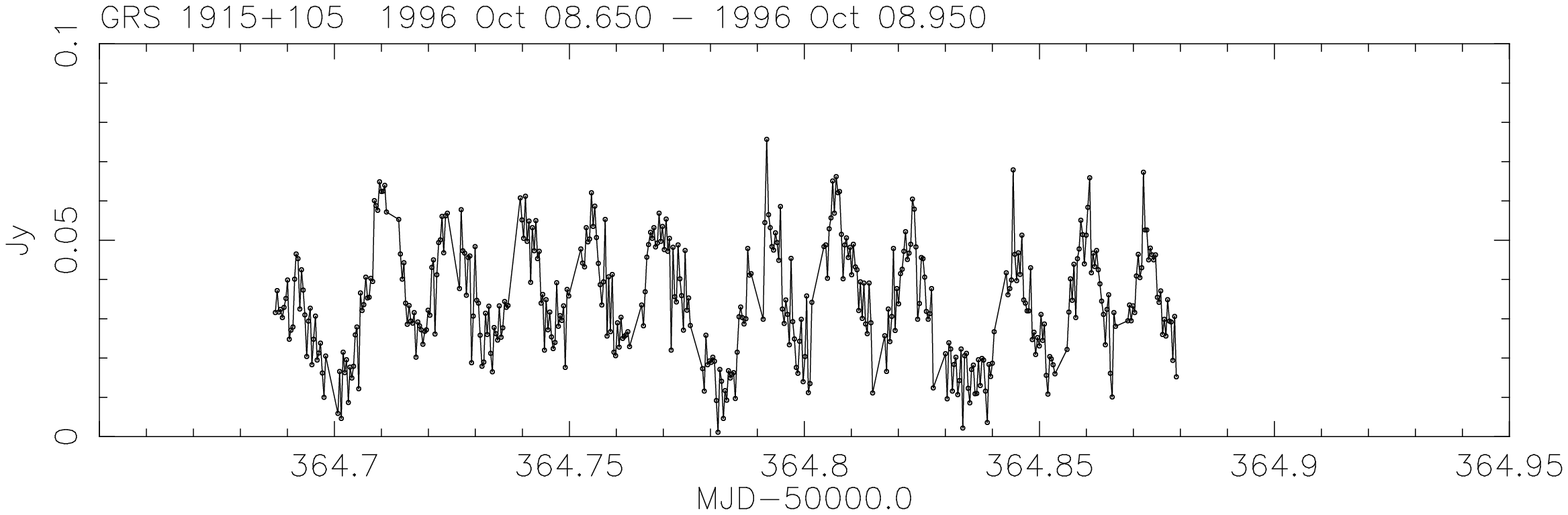,angle=270,width=85mm,clip=}\quad\epsfig{file=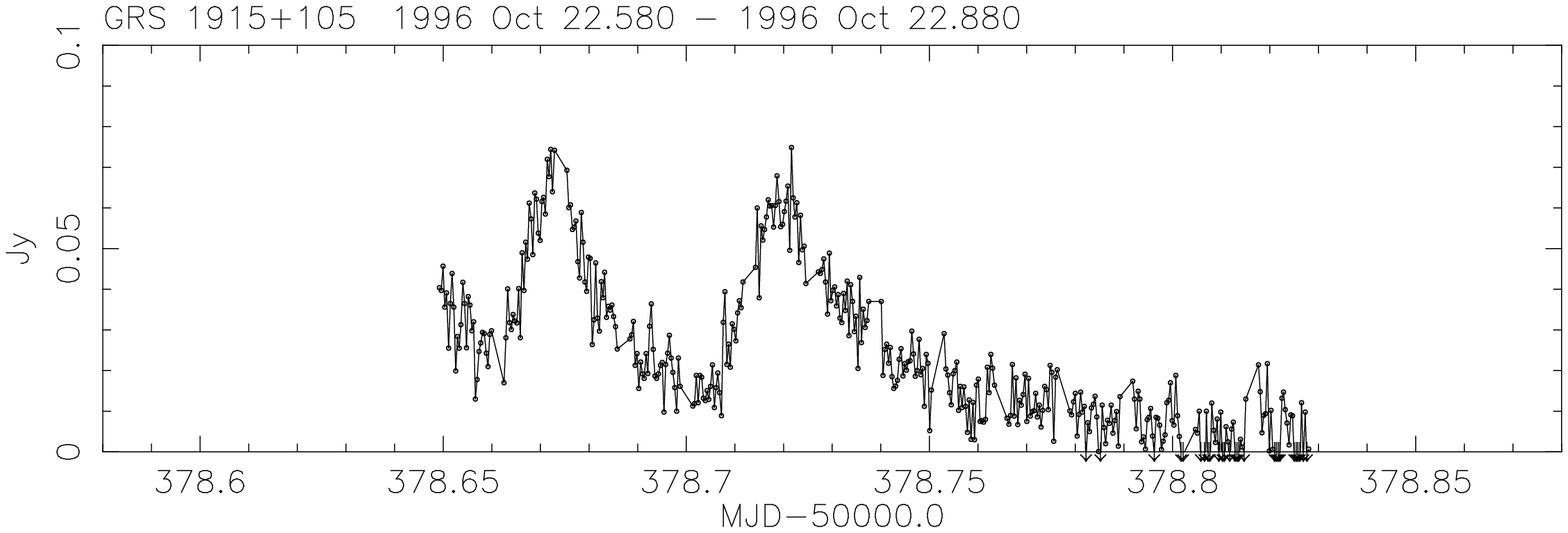,angle=270,width=85mm,clip=}}
\vspace{1mm}
\centerline{\epsfig{file=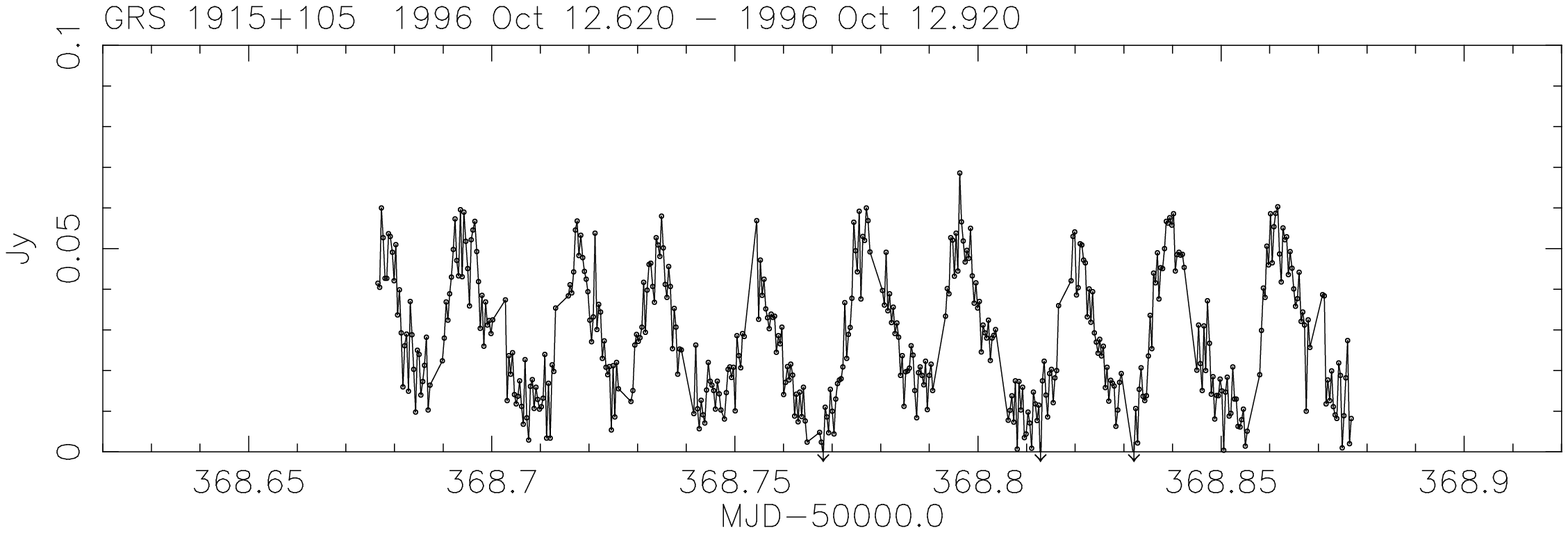,angle=270,width=85mm,clip=}\quad\epsfig{file=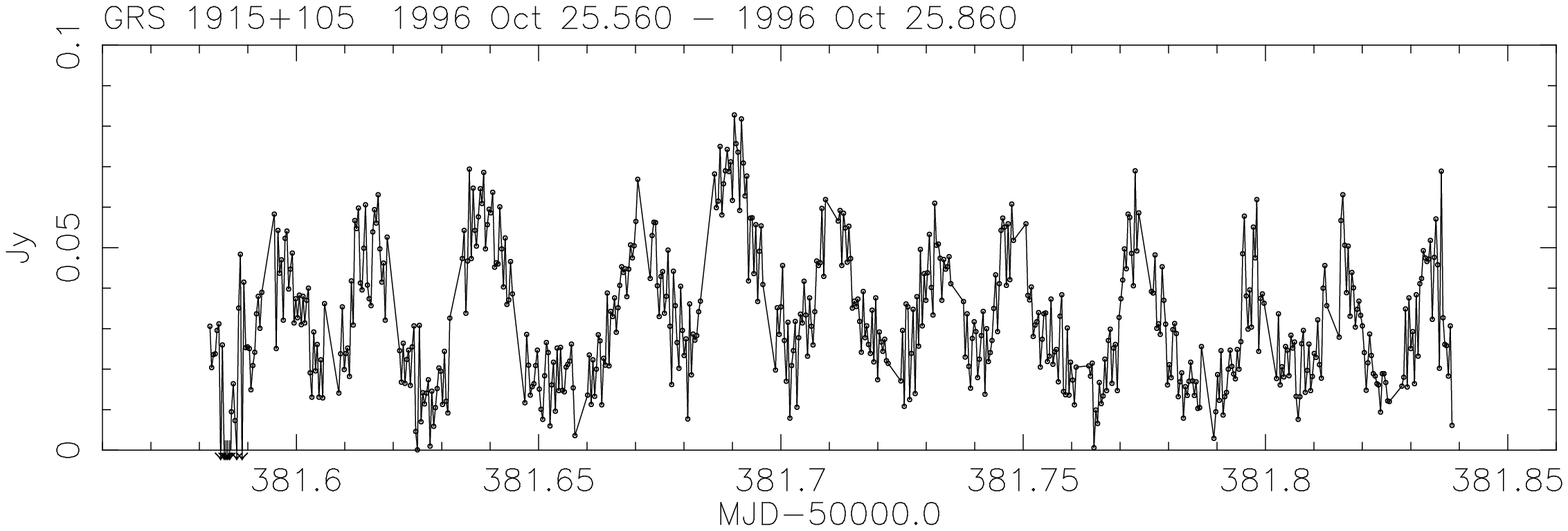,angle=270,width=85mm,clip=}}
\vspace{1mm}
\centerline{\epsfig{file=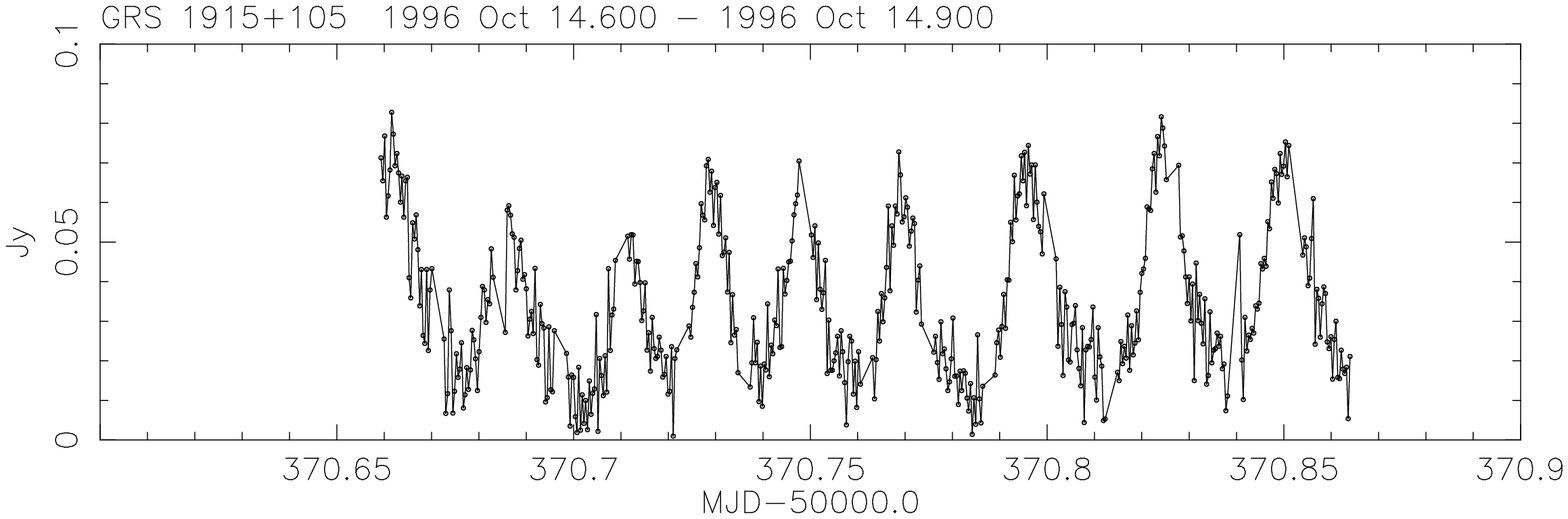,angle=270,width=85mm,clip=}\quad\epsfig{file=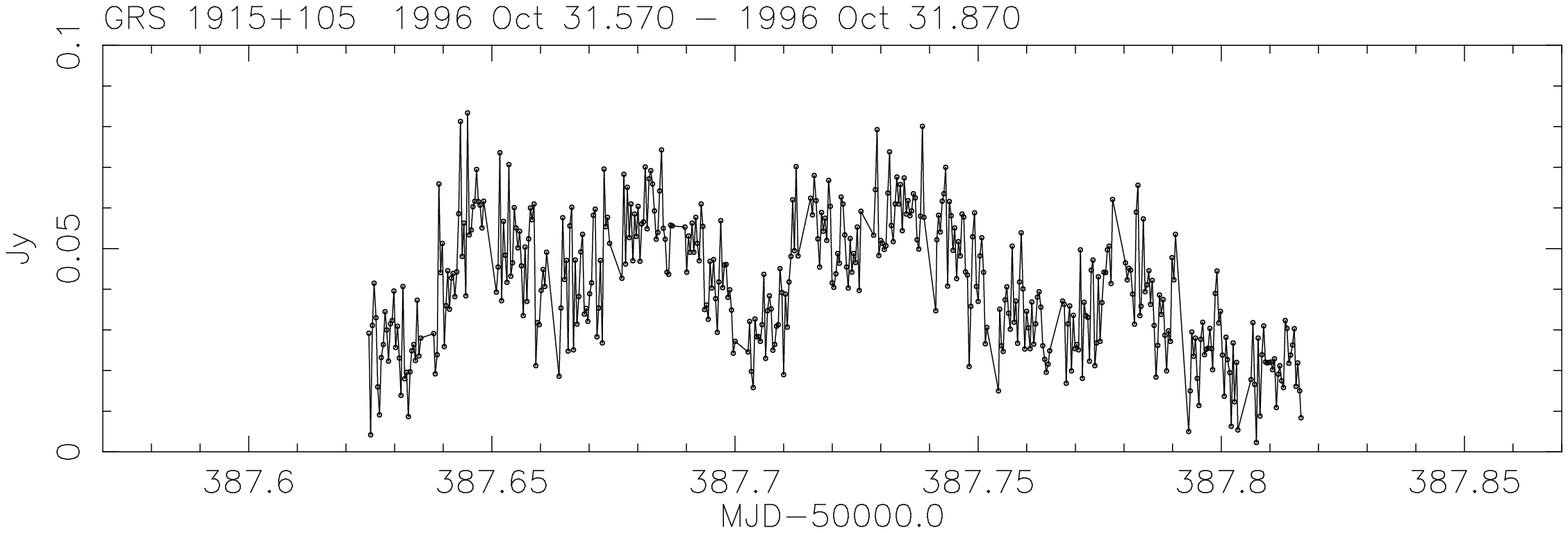,angle=270,width=85mm,clip=}}
\vspace{1mm}
\centerline{\epsfig{file=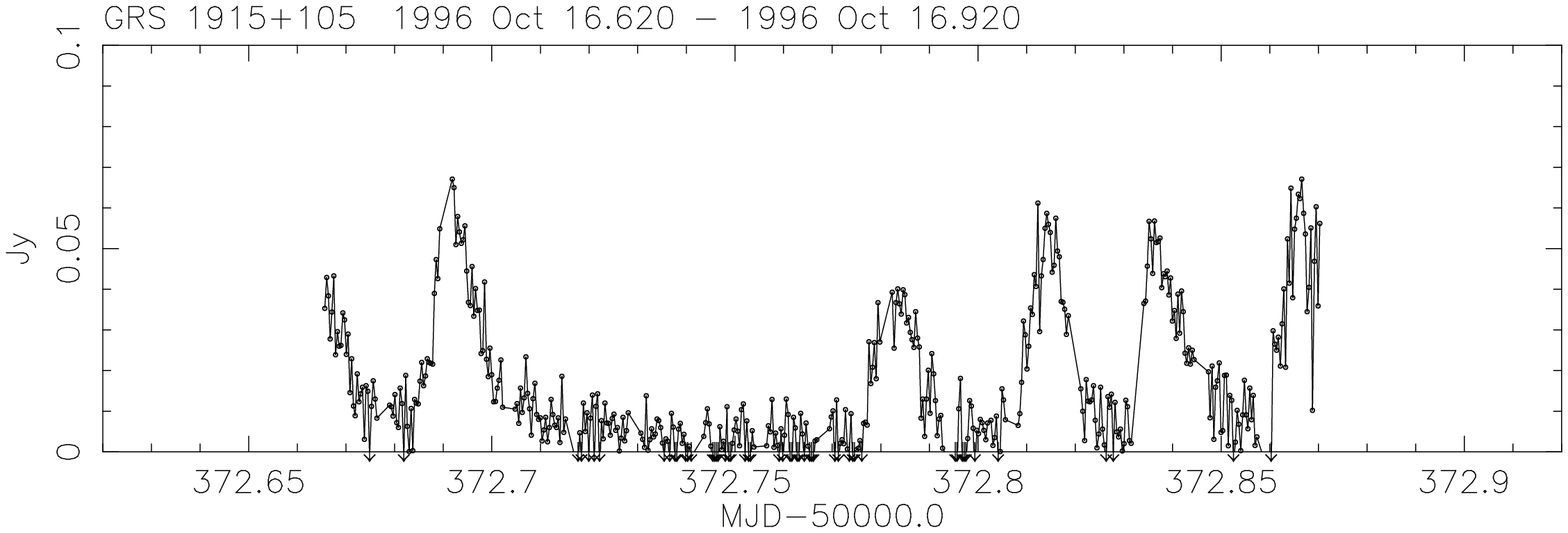,angle=270,width=85mm,clip=}\quad\epsfig{file=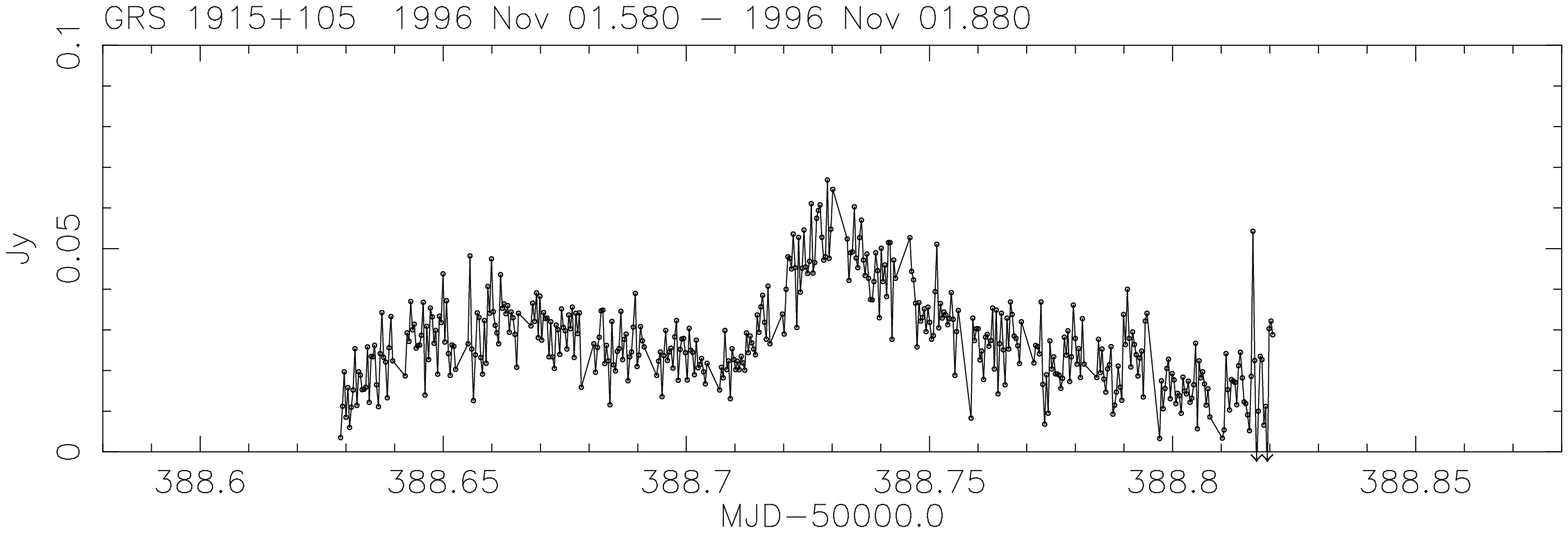,angle=270,width=85mm,clip=}}
\vspace{1mm}
\caption{Short-timescale behaviour of GRS\,1915+105 at 15 GHz; the integration time is 32 s}
\end{figure*}

2. The source seems to have `active' and `passive' periods, with
abrupt changes of state between the two. The most prominent example is
the sequence of 4 observations on 1996 May 23 -- 26 which
showed strong QPO activity, but which were surrounded by long passive
periods. The abrupt changes sometimes last for about 1 day (e.g. 1996
Sep - Nov, see Fig. 4), and in a few cases the change has been
recorded (1996 Sep 16 and 17).

3. The event of 1996 May 24 was observed simultaneously with the VLBA
at 8.3 GHz.
The two sets of data, over the 2 hours of overlap (limited by the
longitude difference between the instruments), are plotted in Fig.
3. There is a delay, in the sense that the 15-GHz data lead the
8.3-GHz, of about 4 -- 5 minutes (based on a subjective estimate of the
best fit when sliding one relative to the other). There are also
substantial changes in spectral index, as can be seen from an
inspection of Fig. 3.
The flux density observed by the VLBA at 2.3 GHz during the same run
was nearly constant, a result consistent with a model in which
lower-frequency emission
comes from a larger photosphere, presumably
determined by optical-depth effects, and with these emission regions being excited
with differing delays. For frequencies where the photosphere exceeds
10 or 20 light-minutes in size, only slow variations would be seen. The
increase in the amplitude of variations with increasing frequency is
confirmed by inspection of the 2.25 and 8.3-GHz data from the
Greenbank interferometer for the flare in 1995 Aug (Foster et al.
1996: see their Fig. 3). Our 15-GHz data on 1995 Aug 11 (Fig. 2), which
have only a very short overlap with the Greenbank data, show larger
variations.

\begin{figure}
\centering
\leavevmode\epsfig{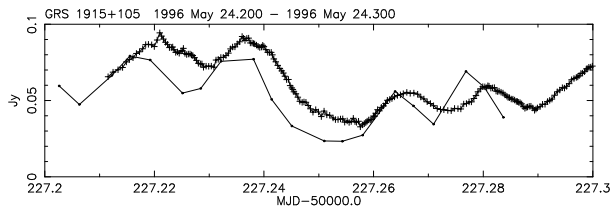}
\caption{Oscillations on 1996 May 24, observed simultaneously with Ryle
Telescope at 15 GHz (dots) and the six shortest baselines of the VLBA at 8.3 GHz (crosses).}
\end{figure}

\begin{table}
\caption{RXTE pointed observations coinciding with RT data}
\begin{tabular}{@{}cccl}

date     &   RXTE times  &  RT times    &  RT data \\
1996 Jun 12 & 0141--0200  & 0051--0337   &   no detection\\
1996 Jun 12 & 0317--0336  & & \\
1996 Jun 25 & 0154--0344  & 0006--0445   &   no detection\\
1996 Sep 07 & 1803--2309  & 1712--2320   &   Fig. 5a\\
1996 Sep 25 & 1600--1829  & 1556--1855   &   marginal detection\\
1996 Oct 15 & 1509--2206  & 1537--2051   &   no detection\\
1996 Oct 26 & 1152--1806  & 1530--1830   &   QPO: Fig. 5b\\
1996 Oct 29 & 1155--1728  & 1637--1956   &   marginal detection\\
1996 Nov 14 & 1505--1740  & 1334--1838   &   20 mJy for 30 min\\
1996 Dec 31 & 0648--1018  & 0919--1219   &   marginal detection\\
\end{tabular}
\end{table}

The 1996 May 24 event  is the only one for which dual-frequency data are
available; the RT is a single-frequency instrument, and the unpredictability of
the source makes it difficult to repeat the observation -- but we regard this as
important.

\setcounter{figure}{3}
\begin{figure*}
\centering
\centerline{\epsfig{file=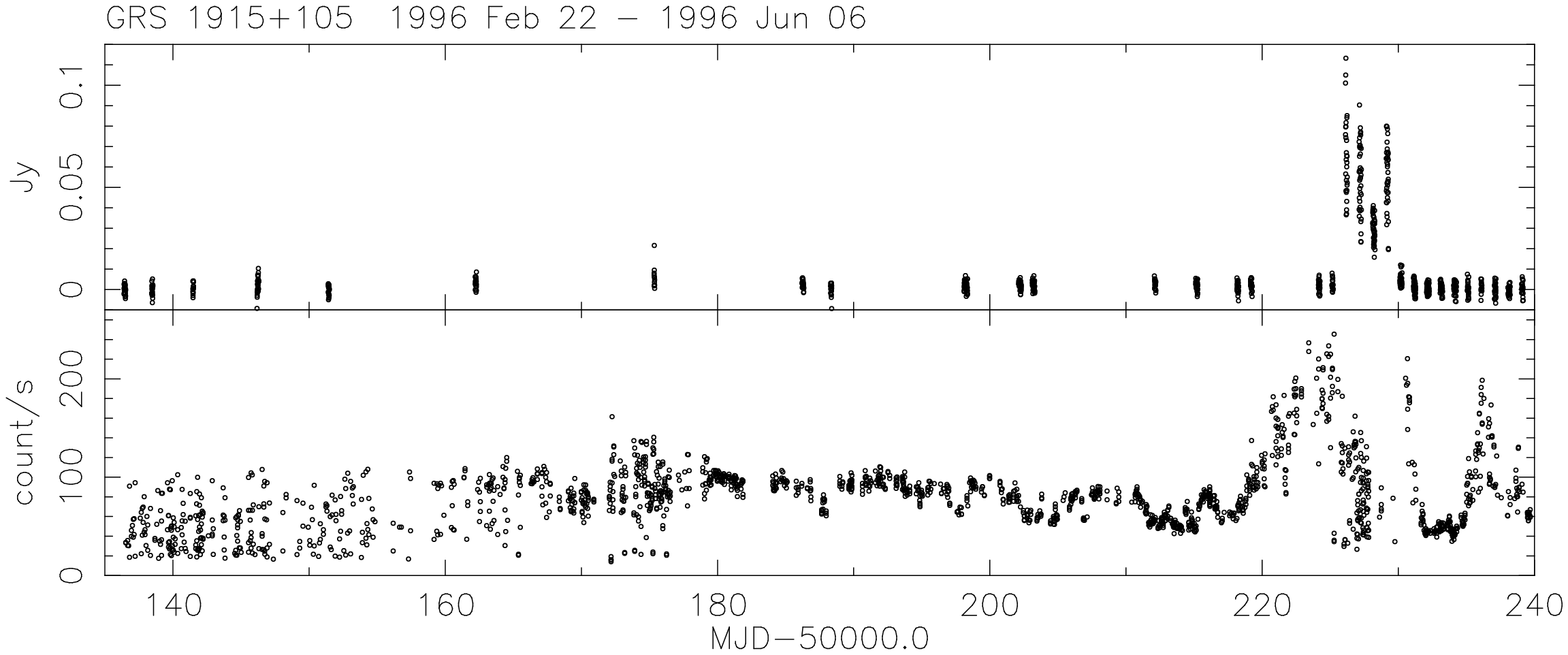,angle=270,width=17cm,clip=}}
\vspace{5mm}
\centerline{\epsfig{file=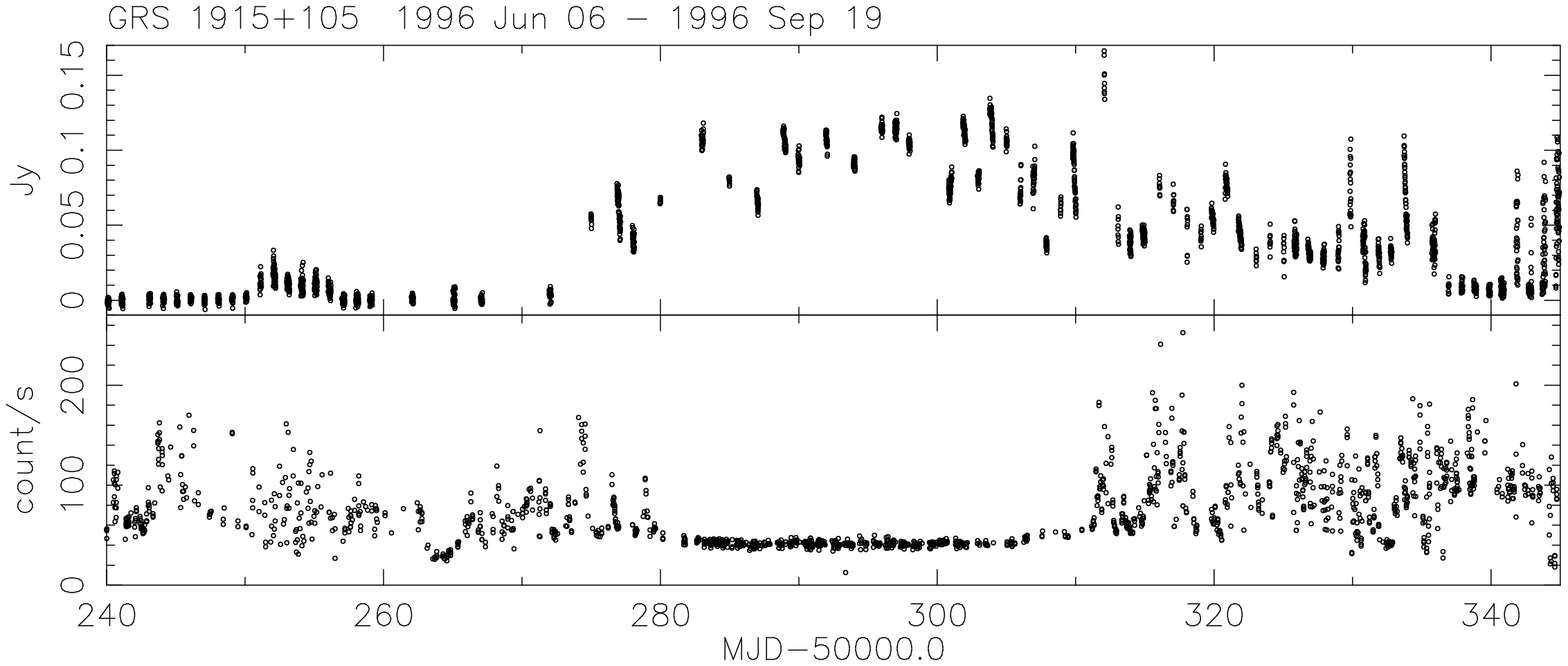,angle=270,width=17cm,clip=}}
\vspace{5mm}
\centerline{\epsfig{file=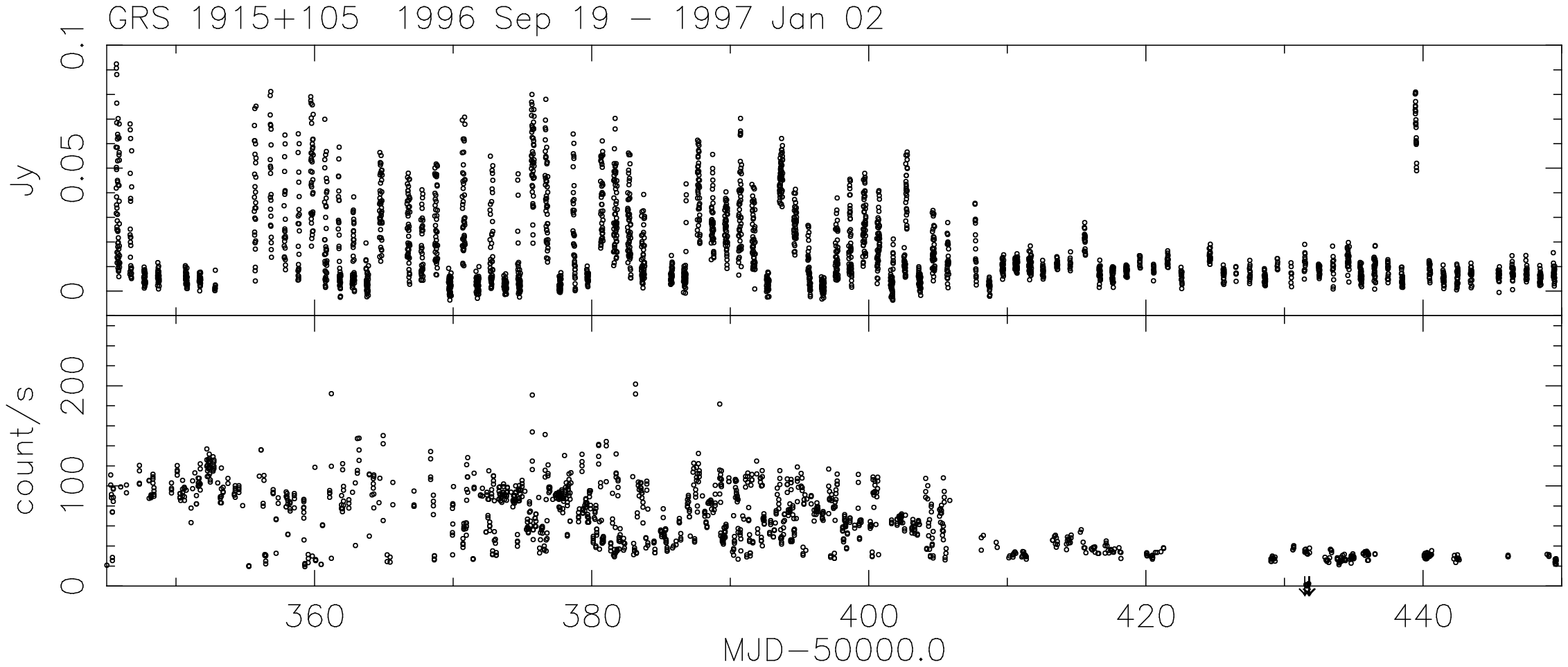,angle=270,width=17cm,clip=}}
\caption {Ryle Telescope (upper panels) and {\it RXTE} ASM observations
(lower) of GRS\,1915+105 from 1996 Feb 22 to 1997 Jan 02}
\end{figure*}

\setcounter{figure}{4}

\begin{figure*}
\centering
\centerline{\epsfig{file=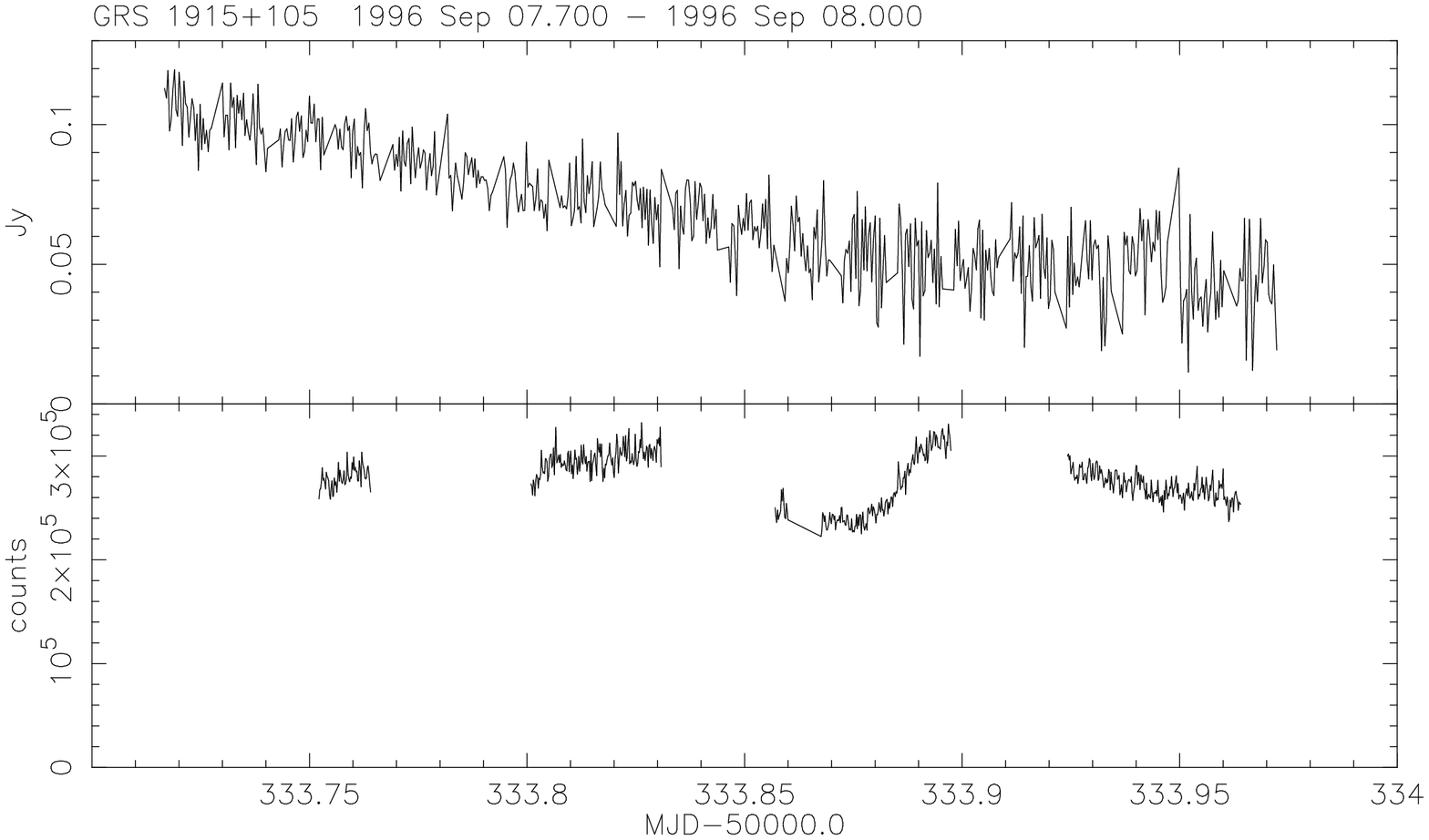,angle=270,width=16cm,clip=}}
\vspace{5mm}
\centerline{\epsfig{file=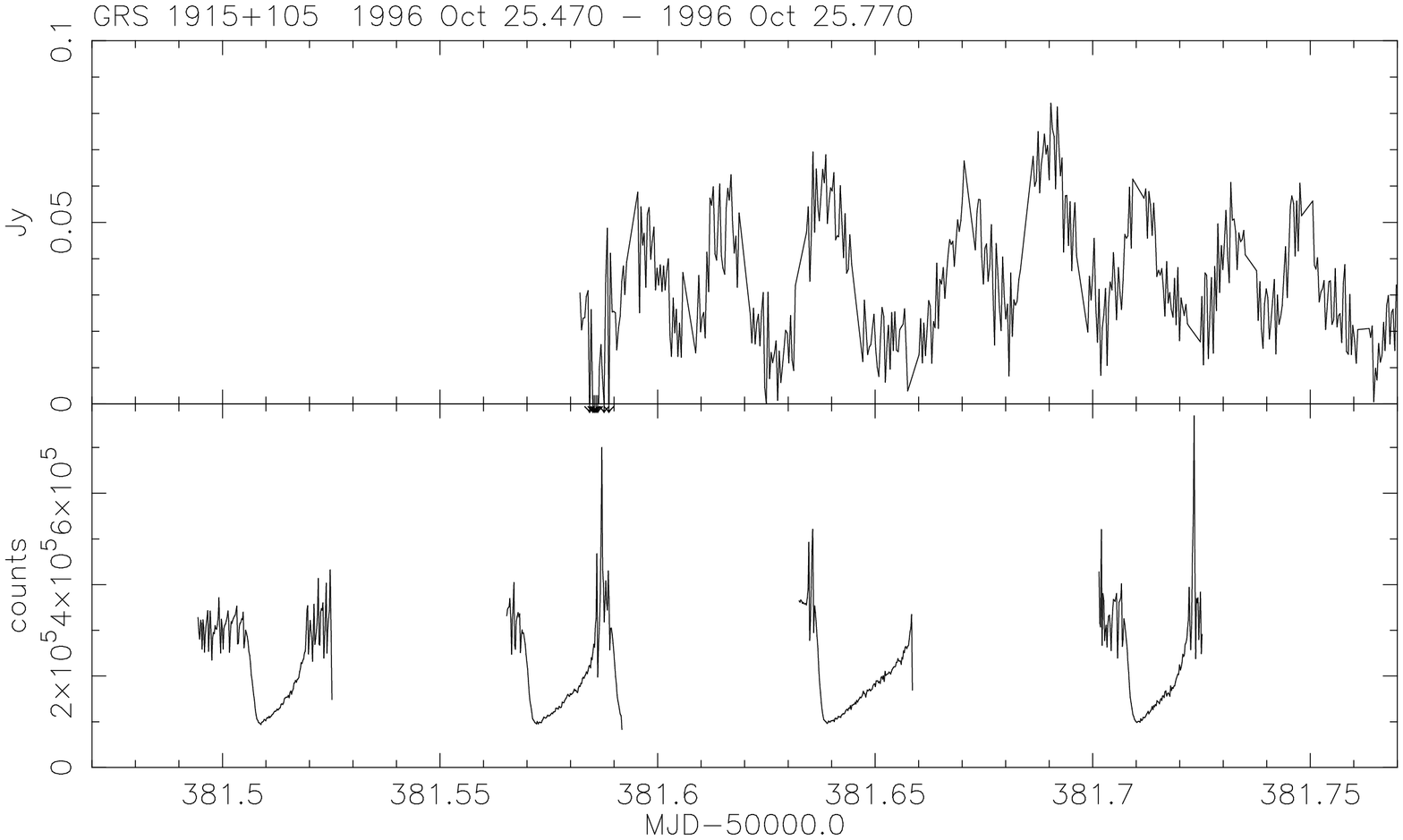,angle=270,width=16cm,clip=}}
\vspace{5mm}
\caption{Ryle Telescope (upper) and {\it RXTE} PCA (lower) observations on 1996 Sep 07 and Oct 25.
The PCA data are in 16-s bins. }
\end{figure*}

\subsubsection{Comparison with soft X-ray {\it RXTE} ASM data}

As a first attempt to consider the X-ray/radio correlations, we have used the
`quick-look' data from the {\it RXTE} ASM experiment, which give integrated
count-rates over the 2 - 10 keV band.

Fig. 4 shows the ASM and 15-GHz data plotted over the same time ranges,
from 1996 Feb 23, when the major part of the ASM dataset starts.

The relationship is complex, and changes through the period of the observations.

MJD 50135 -- 50220: varying degrees of X-ray activity are displayed.
No significant radio emission was detected (although the coverage was rather
sparse) apart from one event on MJD 50146, when a 10-mJy event lasting only
10 minutes was observed. This occured during a period of somewhat enhanced
X-ray activity.

MJD 50226 -- 50229 (1996 May 23 -- 26): radio emission was observed on each of
these 4 days (all are displayed in Fig. 2). They had been preceded by 5 days of
enhanced X-ray actvity: the X-ray emission during the radio flare was lower on average
than in the preceding days, but still appears highly variable.

MJD 50251 -- 50260: a minor radio flare, about 20 mJy, which was also associated
with a fall in an otherwise active X-ray state.

MJD 50275 - 50310: a major radio flare, with no detected QPOs. The examples of
the smooth variation of emission shown in Fig. 2 during this flare are typical.
This flare was observed at a number of radio observatories; results will be
presented elsewhere. Shortly after the start of the radio flare, the X-ray
emission became very steady.

MJD 50310 -- 50340: the radio flare faded (but not monotonically).
QPOs detected on MJD 50329 (Fig. 2), 50335.

MJD 50340 -- 50410: strong QPOs observed on most days, but with interspersed
periods of barely detectable flux density. See Fig. 2 for examples.

MJD 50411 -- 50450: the X-ray flux fell to low levels at approximately the same
time as the radio. There was almost no detectable radio emission, apart from
one day (MJD 50439); a similar event was also reported on MJD 50444 using
the Greenbank Interferometer (E. Waltman, private communication); it was missed
by the RT because of maintenance work.

\subsubsection{Comparison with pointed {\it RXTE} data}

We have searched for observations which coincided with pointed {\it RXTE}
observations.  There are 10 of these; they are detailed in table 1.

Only two of these observations have sufficient radio flux densities to
make useful comparisons. On 1996 Sep 7 (Fig. 5(a)), the radio emission
varied smoothly and the X-ray emission at this time-resolution was
relatively featureless. 1996 October 25 shows the only coincident radio
and X-ray QPOs so far recorded (Fig. 5(b)).  The high X-ray count-rates
coincide with the times of low radio emission. More observations are needed
to allow comparison over all phases, but this result clearly indicates that
the same mechanism drives the two oscillations. On the other hand, the presence
of X-ray oscillations, as for example on 1996 Oct 15, does not imply that
strong radio oscillations will be observed -- there was little detectable radio
emission at that time.

\section{Conclusions}

A wide range of previously unknown phenomena have been recorded in the
high-frequency radio emission from GRS\,1915+105. A strong link has been
established between the radio and X-ray emission.
Fender et al.\ (1997) also report emission in the infrared varying
on similar time-scales to those of the radio oscillations reported here.
They present evidence that each oscillation is associated with an ejection
event, and interpret the infrared flux as the high-frequency tail
of a synchrotron spectrum.
 Belloni et al. (1997) have shown that the soft-X-ray dips on
timescales near 30 min can well be explained by the removal of the inner
200 km of the accretion disc. The coincidence of the rise in the radio
flux density with the X-ray dip in Fig. 5(b) therefore suggests that at
least part of the inner disc is ejected from the system during the
oscillations. Important
observations remain to be made, including simultaneous observations
with high time resolution at as many frequencies as possible.
We suggest also that the major radio outburst starting near MJD 50275,
and the simultaneous nearly-constant X-ray flux, may have been the results
of the removal of a larger part of the inner accretion disc.

\section*{Acknowledgements}

We thank the staff of MRAO for the maintenance and operation of the Ryle
Telescope, whch is supported by the UK Particle Physics and Astronomy
Research Council. We thank Vivek Dhawan for supplying the 8-GHz data from the
VLBA, and Edward Morgan for help with extracting the {\it RXTE} pointed observations.
The ASM  quick-look data are provided by the ASM/{\it RXTE} teams at MIT and at the
{\it RXTE} SOF and GOF at NASA's Goddard Space Flight Center.

\end{document}